\documentclass[11pt]{article}
\usepackage{moriond,epsfig}
\usepackage{wrapfig}

\newcommand{\etall}{{\itshape et al.}}
\newcommand{\journal}[4]{{\em #1}$\;${\bf #2}(#3)#4}

\newcommand{\prl}{\journal {Phys. Rev. Lett.}}

\newcommand{\pl}{\journal {Phys. Lett.}}

\newcommand{\apj}{\journal {Astrophys. J.}}
\newcommand{\ap}{\journal {Astropart. Phys.}}

\newcommand{\nature}{\journal {Nature}}

\newcommand{\PR}{\journal {Phys. Rev.}}

\newcommand{\nim}{\journal {Nucl. Instr. Meth.}}
\newcommand{\rpp}{\journal {Rep. Prog. Phys.}}

\def\be{\begin{equation}}
\def\ee{\end{equation}}
\def\bc{\begin{center}}
\def\ec{\end{center}}
\def\bea{\begin{eqnarray}}
\def\eea{\end{eqnarray}}

\def\amanda{{AMANDA}}
\def\icecube{{Icecube}}
\def\batse{{BATSE}}

\def\edelweiss{{Edelweiss}}

\newcommand{\nutau}{\ensuremath{\nu_{\tau}}}
\newcommand{\nue}{\ensuremath{\nu_{{\mathrm e}}}}
\newcommand{\numu}{\ensuremath{\nu_{\mu}}}

\newcommand{\gev}{\ensuremath{{\mathrm{GeV}}}}
\newcommand{\tev}{\ensuremath{{\mathrm{TeV}}}}
\newcommand{\pev}{\ensuremath{{\mathrm{PeV}}}}
\newcommand{\eev}{\ensuremath{{\mathrm{EeV}}}}

\newcommand{\is}{\ensuremath{{\mathrm{s}^{-1}}}}
\newcommand{\isr}{\ensuremath{{\mathrm{sr}^{-1}}}}
\newcommand{\icm}{\ensuremath{{\mathrm{cm}^{-2}}}}
\newcommand{\units}{\gev\,\icm\,\is\,\isr}

\begin{document}
\vspace*{4cm}
\title{\bc AMANDA: STATUS AND LATEST RESULTS \ec}

\bc
\author{Mathieu Ribordy\\ 
}
\address{{\it Universit\'e de Mons-Hainaut, 19, av. Maistriau, B-7000 Mons}\\
for the \amanda\, collaboration \ec
\vspace*{-0cm}
{\small\begin{sloppypar}                                                                                                         
\noindent
M.~Ackermann$^{4}$,
J.~Ahrens$^{11}$, 
H.~Albrecht$^{4}$,
X.~Bai$^{1}$, 
R.~Bay$^{9}$,
M.~Bartelt$^{2}$,
S.W.~Barwick$^{10}$, 
T.~Becka$^{11}$, 
K.H.~Becker$^{2}$,
J.K.~Becker$^{2}$
% ...
\end{sloppypar}
                                                                                                         
\vspace*{0.1cm}

{\footnotesize
\begin{sloppypar}                                                                                                         
\noindent
$^{(1)}$Bartol Research Institute, University of Delaware, Newark, DE 19716;
$^{(2)}$Department of Physics, Bergische Universit\"at Wuppertal, D-42097 Wuppertal, Germany;
$^{(3)}$Universit\'e Libre de Bruxelles, Science Faculty CP230, Boulevard du Triomphe, B-1050 Brussels, Belgium;
$^{(4)}$DESY-Zeuthen, D-15735, Zeuthen, Germany;
$^{(5)}$Blackett Laboratory, Imperial College, London SW7 2BW, UK;
$^{(6)}$Dept. of Technology, Kalmar University, S-39182 Kalmar, Sweden;
$^{(7)}$Lawrence Berkeley National Laboratory, Berkeley, CA 94720, USA;
$^{(8)}$Dept. of Physics, Pennsylvania State University, University Park, PA 16802, USA;
$^{(9)}$Dept. of Physics, University of California, Berkeley, CA 94720, USA;
$^{({10})}$Dept. of Physics and Astronomy, University of California, Irvine, CA 92697, USA;
$^{({11})}$Institute of Physics, University of Mainz, Staudinger Weg 7, D-55099 Mainz, Germany;
$^{({12})}$Dept. of Physics, University of Maryland, College Park, MD 20742, USA;
$^{({13})}$University of Mons-Hainaut, 7000 Mons, Belgium;
$^{({14})}$Departamento de F\'{\i}sica, Universidad Sim\'on Bol\'{\i}var, Caracas, 1080, Venezuela;
$^{({15})}$Dept. of Physics, University of Wisconsin, Madison, WI 53706, USA;
$^{({16})}$Physics Dept., University of Wisconsin, River Falls, WI 54022, USA;
$^{({17})}$Division of High Energy Physics, Uppsala University, S-75121 Uppsala, Sweden;
$^{({18})}$Dept. of Physics, Stockholm University, SE-10691 Stockholm, Sweden;
$^{({19})}$Vrije Universiteit Brussel, Dienst ELEM, B-1050 Brussels, Belgium;
\end{sloppypar}
}
}
}

\maketitle\abstracts{
We briefly review some of the recent \amanda\, results emphasizing the all flavor capabilities of the high energy neutrino telescope, important in the context of equal neutrino mixing from distant sources at Earth. 
Together with a report on a preliminary UHE neutrino flux limit, the course of our progress in the quest for point sources is described.
Finally, a 1 year preliminary limit of \amanda-II to neutralino cold dark matter (CDM) candidates, annihilating in the center of the Sun, for various MSSM parameter choices is presented and discussed.
}

\newpage
\section{Motivations}
\begin{wrapfigure}{r}{6.5cm}
\vspace*{-11mm}
\centering
\mbox{\epsfig{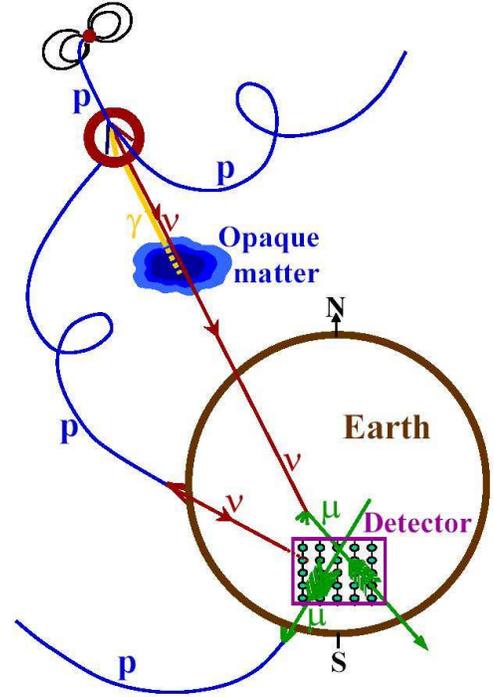}}
\caption{Illustration of the underlying physics performed with \amanda, see text.\hfill~}
\label{fig:principle}
\end{wrapfigure}
The observation of the neutrino sky may eventually shed some light on the obscure origin of the charged cosmic ray (CR) spectrum. The acceleration mechanisms at energies above the knee remain a mystery. CRs are believed to be accelerated in the expanding shocks of e.g. SNR, AGN or GRB~\cite{cr-acc}. Hadrons accelerated in these objects collide with surrounding radiation or 
with background radiation 
between the source and the Earth 
to produce pions, which further decay into neutrinos and gamma rays.
Although it seems to be established that SNRs accelerate electrons
up to $\approx\! 100\,\tev$ through the study of multiwavelength spectra in conventional astronomy, it has not yet been unambiguously demonstrated that they might also accelerate hadrons.
The nascent field of neutrino astronomy may settle the controversy. 
The neutrino travels unabsorbed, undeflected and escapes optically thick sources and so offers an advantage over other astrophysical messengers
(neutrinos are nevertheless difficult to detect and require large detection volumes).
CR direction is randomized by magnetic fields. Gamma rays interact with the 
CMBR and the IR background, which reduces their mean free path above modest energies (10 GeV),
consequently affecting the observed emission spectra of astrophysical objects (the close AGN Mrk501 would be invisible at 100 TeV~\cite{mrk501-spectrum}). 
The neutrino telescope \amanda\, (Antarctic Muon And Neutrino Detector Array) is dedicated to the exploration of the high energy (HE) universe and aims at detecting extraterrestrial HE neutrinos, which may be produced by the powerful processes at work in cosmic accelerators.

\section{The \amanda\, detector}
\begin{wrapfigure}{r}{9.5cm}
\vspace*{-12mm}
\mbox{\epsfig{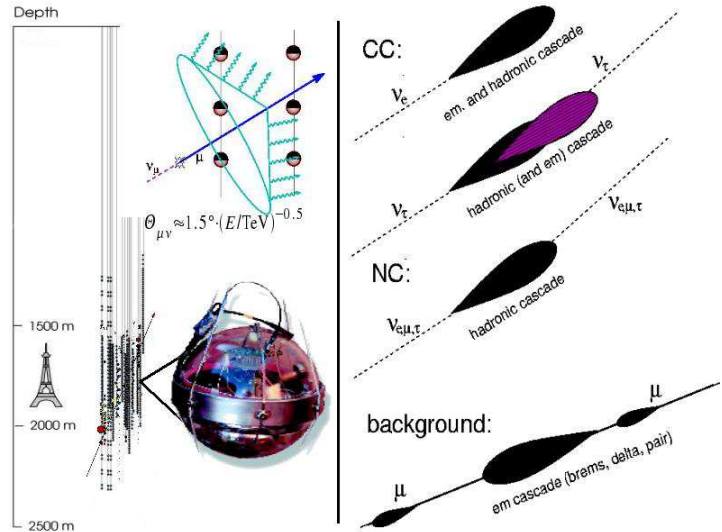}}
\caption{Left: the \amanda\, telescope and the detection principle of muon tracks. Right: ''cascades'' allow for all flavor detection.\hfill~}
\label{fig:principle}
\end{wrapfigure}

AMANDA is buried in the transparent ice of the South Pole ice cap at depths between 1.5 and 2.2 km~\cite{nature}, see Fig.~\ref{fig:principle}, in order to 
escape 
a large fraction of the atmospheric muon flux.
The remaining events, misreconstructed below the horizon, constitute a main source of background.
Atmospheric neutrinos~\cite{atm-nu} represent another important source of background in a astrophysical search.
\amanda\, is a 19 string detector equipped with 677 PMTs instrumenting a cylindrical volume with an outer radius of 60~m.
Events are reconstructed by measuring the arrival time of Cherenkov light emitted either by crossing relativistic (neutrino-induced) muons or by EM/hadronic cascades, which results from NC for all three flavors and $\nu_{\mathrm{e},\,\tau}$ CC interaction~\cite{amanda-reco}.
\amanda\, is therefore an all flavor detector, important in the context of neutrino oscillation, as all flavors are expected to be equally populated at Earth after traveling cosmological distances (given 
\nue:\numu:\nutau=1:2:0 at the source, this relies on the $U_{\mathrm{e}1}\!\approx\!0$ and $U_{\mu3}\!\approx\! U_{\tau3}$ in the neutrino mixing matrix~\cite{cosmnumixing}).
Muon tracks are reconstructed with an incidence direction resolution $\Delta\Psi\!\approx\! 2.5^\circ$, cascades however have a better energy resolution $\Delta\mathrm{log}E\!\approx\! 15\%$. 
The \amanda\, analyses have to face various systematical errors, which come from an imperfect knowledge of the ice properties, from uncertainties in the absolute detector sensitivity and in the primary CR spectrum normalization and its exact composition.

\section{Diffuse flux analyses}
This section discusses analyses aimed at finding global neutrino excess contributed to for example by unresolved sources of a possible cosmogical origin.
They exploit an expected harder extra-terrestrial spectrum ($d\Phi/dE\!\!\sim\!\! E^{-\alpha}, \alpha\!\!\approx\!\! 2$) in comparison to the steeply falling atmospheric neutrino background ($\alpha\! =\! 3.7$). 
These analyses critically depend on the quality of the detector simulation and can be performed in the two complementary muon and cascade channels.

\subsection{Search for an UHE neutrino excess}\label{subsec:uhe}
Above $\sim\,\!\! 1\pev$, because of the rise of the neutrino cross section with increasing energy, the Earth becomes gradually opaque and neutrino events concentrate near the horizon.
An analysis of the \amanda-B10 '97 data was conducted focusing on PeV to EeV energies~\cite{hundertmark}.
In this range, a crossing muon illuminates the whole detector so different event selection techniques had to be developed (Ref.~\cite{diffuse97}).
Discriminant observables for this analysis used to distinguish UHE neutrino-induced muon from atmospheric muon bundles are: the fraction of hit channels with exactly one hit, the number of hits, the number of hit channels, the averaged amplitude of the hit channels and the reconstruction quality. At these energies, new sources of systematical error caused by the uncertainties on the neutrino cross section and on the muon propagation are taken into account.
Given a $E^{-2}$ benchmark neutrino spectral shape, a preliminary limit of $E^2 \Phi_\nu(E)\!<\! 1.5 \cdot 10^{-6} \,\units$ in the range $1\,\pev\!<\!E\!<\!3\,\eev$ is set, assuming \nue:\numu:\nutau=1:1:1 and $\nu/\bar\nu=1$.

\subsection{Cascade analysis 2000}
\begin{wrapfigure}{r}{6.5cm}
\vspace*{-12mm}
\mbox{\epsfig{file=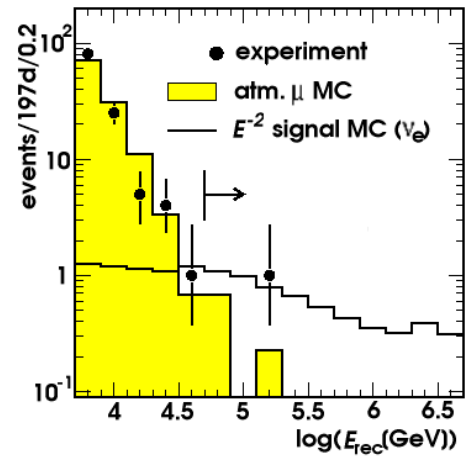,angle=0,width=0.40\textwidth}}
\caption{The experimentally reconstructed energy distribution after the final selection agrees with the expected background. The corresponding response to an hypothetical signal is also shown.\hfill~}
\label{fig:cascadecut}
\end{wrapfigure}
\begin{figure}
\begin{minipage}[t]{0.4\linewidth}
\begin{center}
\includegraphics[width=6cm,clip]{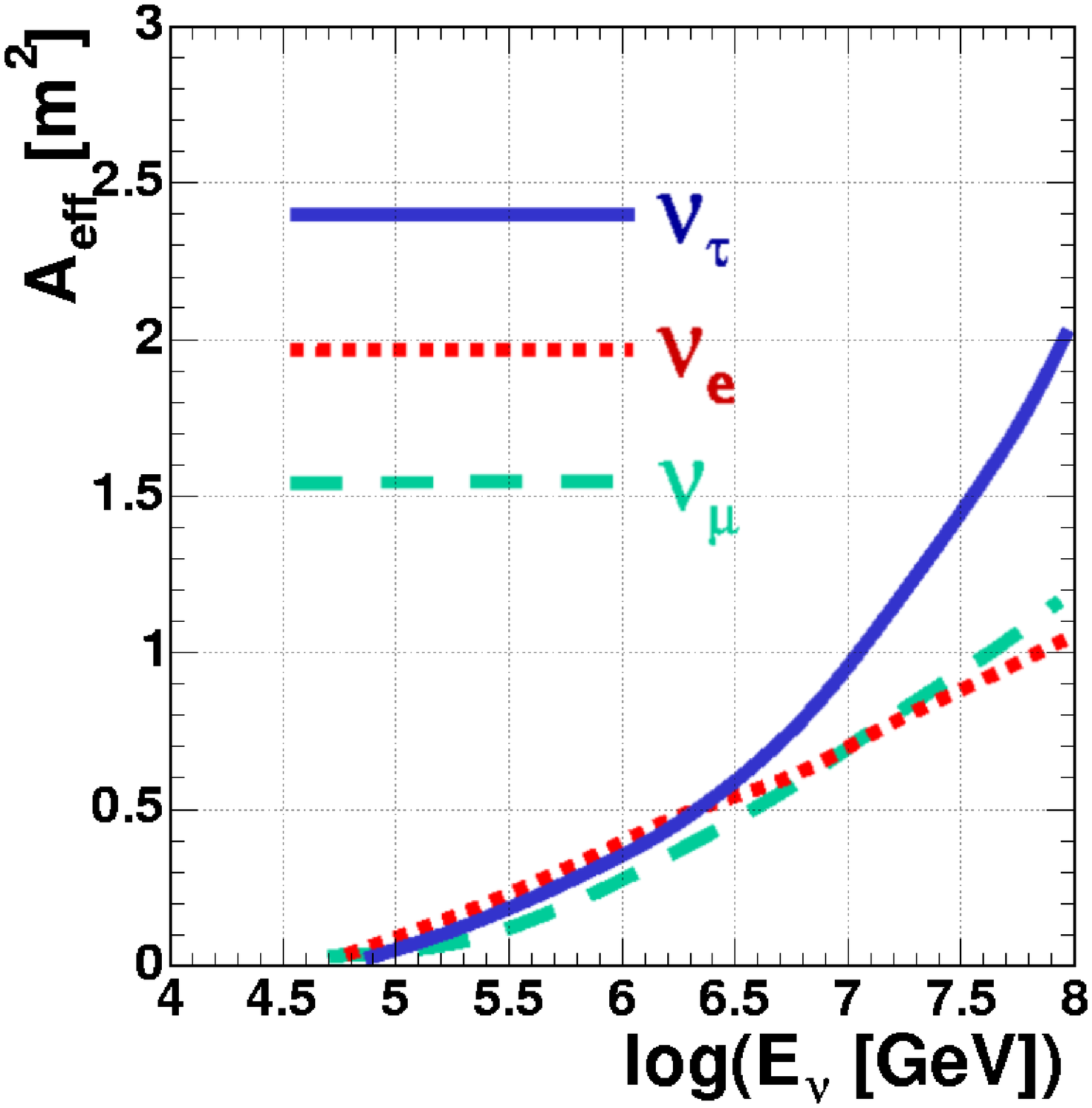}
\caption{\label{fig:cascadeaeff}The effective area w.r.t. the energy allows to compute a limit for any model (arbitrary spectrum and neutrino mixing).\hfill~}
\end{center}
\end{minipage}
\hfill
\begin{minipage}[t]{0.55\linewidth}
\begin{center}
\includegraphics[width=8cm,clip]{figures/neutrino-diffuse-flux.ps2}
\caption{\label{fig:diffuselimit}Summary of the \amanda\, and \icecube\, diffuse flux limits in red (plain: published, dotted: ongoing analysis and expected in the future). See also text.\hfill~ }
\end{center}
\end{minipage}
\end{figure}

Strategies followed in this analysis are inspired 
by the reconstruction and selection techniques originating
in a previous study~\cite{cascade97}. These have now been further developed.
Neutrinos from each flavor were generated,
as well as a massive amount of atmospheric muon background (2.5 yr).
An optimized sequential selection was applied to the experimental and to the simulated data in order to reject the atmospheric muon background, reduced by a factor better than $10^{9}$, while preserving the efficiency on the signal, close to 3\% for \nue. The selection 
restricted the reconstructed threshold energy $E_\mathrm{rec}>50\,\tev$ and the reconstruction quality ${\cal L}_{E_\mathrm{rec}}(E)$, demanding the cascade to be contained. 
Some of the restrained observables included among others:
the smoothness (a measure of the equirepartition of hits along the track), the number of hit channels, the number of direct photons and the radial distance between two independently reconstructed vertices (using two complementary sets of hit channels).
One event remained in the final sample of experimental data, with an energy $E_\mathrm{rec}\!\approx\! 150\,\tev$. 
This is shown in Fig.~\ref{fig:cascadecut}. 
Within the estimated systematical uncertainties,
the energy distribution is in agreement with the simulated background.
Obtained sensitivities for all flavors are comparable (Fig.~\ref{fig:cascadeaeff}),
demonstrating the \amanda\, detector as an all flavor neutrino telescope. 
The upper limit reached in this analysis is $E^2 \Phi_\nu(E)  < 8.6 \cdot 10^{-7} \,\units$ in the range $50\,\tev\!<\!E\!<\!5\,\pev$, assuming \nue:\numu:\nutau=1:1:1 and a $E^{-2}$ neutrino spectrum. Some of the SDSS models~\cite{sdss,ssq} are therefore discarded. 
A sensitivity at the same level is foreseen in an on-going analysis in the muon channel. 
Potential, preliminary and published limits are summarized in Fig.~\ref{fig:diffuselimit}, also indicating the level of the atmospheric neutrino flux (\numu\, and \nutau)~\cite{lipariatmtau}, of the cosmogenic flux~\cite{cosmogenic}, the WB and MPR upper limits~\cite{wb,mpr} and a specific flux prediction (MPR~\cite{mpr}).

\section{Point source analyses}
This section briefly illustrates analyses searching for a statistical excess originating in
narrow regions of the northern sky. These analyses exclusively rely on the muon channel, because of its better pointing resolution. The sensitivities of these analyses are optimized by taking advantage of the experimentally observed off-source detector response thus defining the background.

\subsection{Gamma ray burst analysis}
The detection of a HE neutrino component ($E\!>\! 10^{2}\,\tev$) spatially and temporally coinciding with GRBs would substantiate the hypothesis of hadronic acceleration occuring within the GRB wind.
The cumulative \amanda\, data ('97-'00), within a 10 minute time window around 317 \batse\, reported triggered bursts~\cite{batse}, were explored in search of a global excess~\cite{grb}.
The stability of the detector was assessed by evaluating the noise rate within 1 hour from the trigger times, leading to the exclusion of 5 \batse\, triggers.
A very low background analysis (due to the known time-stamp and direction of the burst) with a large muon effective area $A_\mathrm{eff}\approx 50'000 \,\mathrm{m}^2$ (at $E\approx\pev$) was subsequently performed and no events were observed. An event upper limit of 1.45 was derived.
Assuming a Waxman-Bahcall type spectrum, in the GRB fireball phenomenology~\cite{grb-fireball}, a 90\% C.L. upper limit to its normalization constant is set to $4.8 \cdot 10^{-8} \,\units$ ($E_B\!=\! 100\,\tev$ and $\Gamma\!=\! 300$).

\subsection{Summary of the generic 2000 point source quest}
\begin{figure}
\begin{minipage}{0.48\linewidth}
\begin{center}
\includegraphics[width=0.8\textwidth,clip]{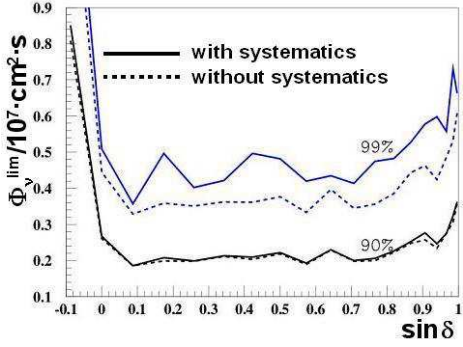}
\caption{\label{fig:ptsource}\amanda-II sensitivity w.r.t. the declination. Not only a quantitative improvement in comparison to \amanda-B10 but also a qualitative improvement on the near horizon sensitivity is achieved.
\hfill~}
\end{center}
\end{minipage}
\hfill
\begin{minipage}{0.48\linewidth}
\label{tab:ptsource}
\begin{center}

\vspace*{1mm}~

\begin{tabular}{|r|c|c|c|}
\hline
{\bf sources} & {\bf declination} & {\bf 1997} & {\bf 2000} \\
\hline
SS433 & 5.0$^\circ$ & - & 0.7 \\
M87 & 12.4$^\circ$ & 17.0 & 1.0 \\
Cas. A & 58.8$^\circ$ & 9.8 & 1.2 \\
Cyg. X-3 & 41.0$^\circ$ & 4.9 & 3.5 \\
Mrk501 & 39.8$^\circ$ & 9.5 & 1.8 \\
Mrk421 & 38.2$^\circ$ & 11.2 & 3.5 \\
Crab & 22.0$^\circ$ & 4.2 & 2.4 \\
\hline
\end{tabular}

\vspace*{9mm}

\parbox{0.95\linewidth}{\footnotesize Table 1: Upper flux limit in $10^{-8}\icm\is$ units reached in the '97 and '00 analyses for some candidates, assuming a $E^{-2}$ spectral shape integrated above $E_\nu=10\,\gev$.~\hfill}
\end{center}
\end{minipage}
\end{figure}

An analysis searching for point sources in the 2000 data, developed by adopting the blindness requirement, as described in Ref.~\cite{ptsource00}, follows an earlier analysis with 1997 data~\cite{ptsource97} and shows a greatly improved sensitivity (Fig.~\ref{fig:ptsource}), particularly near the horizon ($\delta=0$), due to the extended radius of \amanda-II.
699 events have been isolated below the horizon with a purity of about $95\%$ ($sin(\delta)>0.1$), in good agreement with atmospheric neutrino expectations. The integrated declination averaged sensitivity above 10 GeV to point source is $\Phi_\mathrm{\nu}^\mathrm{1yr}<2.3\cdot10^{-8}\icm\is$.
Once 3 years of data (2000-2002) has been combined,
it is estimated to go down to $\Phi_\mathrm{\nu}^\mathrm{3yr}<0.9\cdot10^{-8}\icm\is$.
Two distinct searches have been performed estimating the background from the off-source event input in the corresponding declination band: the first looking for a random statistical excess and the second focusing on specific source candidates.
However, no evidence of an excess has yet been found.
Notably, after one year of data collection, \amanda-II can detect sources emitting at the level of Mrk501 or Mrk421 during their flary period, given a unit $\gamma/\nu$ ratio (accounting for the modified spectral shape in the TeV range~\cite{mrk501-spectrum}). The projected sensitivity $\Phi_\mathrm{\nu}^\mathrm{3yr}$ is at the detection level of the microquasar SS433 for a specific model~\cite{ss433}.

\subsection{The dark matter connection}
\begin{wrapfigure}{r}{8cm}
\vspace*{-12mm}
\mbox{\epsfig{file=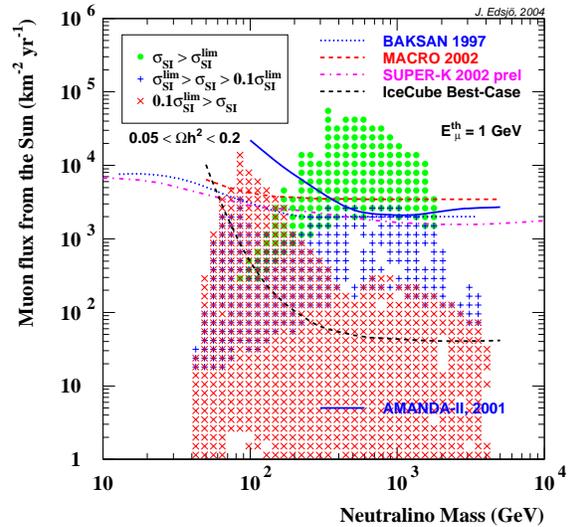,angle=0,width=0.5\textwidth}}
\caption{$\Phi_\mu$ vs $m_\chi$ in \amanda-II for various MSSM models satisfying $0.05\!\!<\!\!\Omega h^2\!\!<\!\! 0.2$. Dots (crosses) are for models which are (not yet) excluded. An \amanda-II preliminary limit is at the edge of unprobed regions.\hfill~}
\label{fig:wimp}
\end{wrapfigure}
AMANDA-B10 performed
an indirect search for non baryonic CDM, in the form of the lightest neutralino $\chi$ in the MSSM framework, from the center of the Earth~\cite{wimp}.
Now, the improved reconstruction capabilities of \amanda-II for nearly horizontal tracks allow a similar search for CDM from the Sun. 
Neutralinos in the halo scattering off ordinary matter may eventually be gravitationally trapped in macroscopic objects. 
Subsequent scattering then reduce their velocity resulting in their accumulation over astronomical times in the core of these objects, where they annihilate pairwise producing neutrinos that can be detected by \amanda.
During 143.7 days of the detector lifetime 2001, the Sun was located below the horizon. An ongoing solar neutralino WIMP analysis has now unblinded the 2001 data, yielding an exclusion limit, shown in Fig.~\ref{fig:wimp}~\cite{edsjo-cdm}, which is at the level of direct CDM search experiments (\edelweiss~\cite{edelweiss-result}), for $m_\chi>500\,\gev$. Once a few years of data taking have been cumulated, the WIMP sensitivity is expected to explore the MSSM parameter space beyond the reach of the current direct CDM search experiment.
It must be stressed that nuclear recoil and indirect CDM search experiments are not equivalent.
In the case of an unevenly distributed CDM halo throughout the galaxy,
the latter have a detection potential which remains intact.
Moreover, solar spin dependant scattering cross section could significantly enhance the trapping rate of neutralinos.

\section{Conclusions and perspectives}
\amanda\, is a neutrino telescope with sensitivity to diffuse fluxes in all three flavor channels, in an energy range extending over 7 orders of magnitude ($\sim\!\!(10^{12},10^{19})\mathrm{eV}$).
It was shown that \amanda-II exhibits a declination averaged sensitivity of $\Phi_\mathrm{\nu}^\mathrm{1yr}<2.3\cdot10^{-8}\icm\is$, greatly improved near the horizon compared to a previous analysis~\cite{ptsource97}.
A search for a response of \amanda-II\, to neutralino annihilation in the Sun is currently being conducted. A preliminary limit was presented, which remarkably suggests that regions of the MSSM parameter space not yet probed will be covered once a few years of data taking have been cumulated.
The upgraded \amanda-II ground hardware of 2003 now has a full digital readout, which allows for single photoelectron resolution and should improve the capabilities of reconstruction for UHE events.
During the next pole season, the first \icecube~\cite{icecube} strings will be installed. The construction of this cubic km neutrino telescope should be completed in 2010. In the time being, the data will be gradually combined with that of \amanda-II, enabling more sensitive analyses and probing the neutrino sky to higher energy.

\vspace*{-1mm}

\section*{Acknowledgments}
\baselineskip=12pt  
{\footnotesize
We acknowledge the support of the following agencies: National
Science Foundation--Office of Polar Programs, National Science
Foundation--Physics Division, University of Wisconsin Alumni Research
Foundation, Department of Energy, and National Energy Research
Scientific Computing Center (supported by the Office of Energy
Research of the Department of Energy), UC-Irvine AENEAS Supercomputer
Facility, USA; Swedish Research Council, Swedish Polar Research
Secretariat, and Knut and Alice Wallenberg Foundation, Sweden; German
Ministry for Education and Research, Deutsche Forschungsgemeinschaft
(DFG), Germany; Fund for Scientific Research (FNRS-FWO), Flanders
Institute to encourage scientific and technological research in
industry (IWT), and Belgian Federal Office for Scientific, Technical
and Cultural affairs (OSTC), Belgium; Fundaci\'{o}n Venezolana de 
Promoci\'{o}n al Investigador (FVPI), Venezuela;  D.F.C. acknowledges 
the support of the NSF CAREER program.
}

\end{document}